# THE STATUS OF TURKIC ACCELERATOR COMPLEX PROPOSAL[*]

S. Sultansoy[#], M. Yilmaz, Gazi University, Ankara, TURKEY,
O. Cakir, A.K. Çiftçi, E. Recepoglu, O. Yavas, Ankara University, Ankara, TURKEY

*Abstract*

Recently, the Turkic Accelerator Complex (TAC) is proposed as a regional facility for accelerator based fundamental and applied research. The complex will include linac on ring type electron-positron collider as a phi, charm and tau factory, linac based free electron laser (FEL), ring based third generation synchrotron radiation (SR) source and a few GeV proton accelerator. Preliminary estimations show that integral luminosity of hundred inverse femto-barns per year can be achieved for factory options. The FEL facility is planned to obtain laser beam between IR and soft X-ray region. In addition, SR facility will produce photon beams in UV and X-ray region. The proton accelerator will give opportunity to produce muon and neutron beams for applied research. The current status of the conceptual study of the complex is presented.

## INTRODUCTION

Particle accelerators technology is one of the generic technologies which is locomotive of the development in almost all fields of science and technology. Because of this, accelerator technology should become widespread all over the world. Existing situation shows that a large portion of the world, namely the South and Mid-East, is poor on the accelerator technology. UNESCO has recognized this deficit and started SESAME project in Mid-East, namely Jordan (see [1] and refs. therein). Turkic Accelerator Complex (TAC) project is more comprehensive and ambitious project, from the point of view of it includes light sources, particle physics experiments and proton and secondary beam applications.

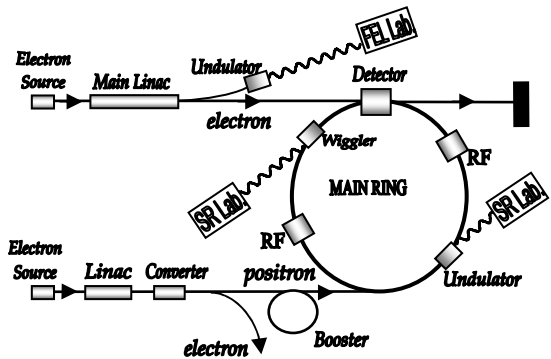

Figure 1: Schematic view of the TAC charm factory complex.

___________________
*Work supported by Turkish State Planning Organisation (DPT)
#saleh@gazi.edu.tr

Approximately 10 years ago, linac-ring type charm-tau factory with synchrotron light source was proposed as a regional project for elementary particle physics [2]. Starting from 1997, a small group from Ankara and Gazi Universities begins a feasibility study for the possible accelerator complex in Turkey with the support of Turkish State Planning Organization (DPT) [3]. The results of the study is published in [4] and presented at EPACs [5-6]. Starting from 2002, the conceptual design study of the TAC project has started with a relatively enlarged group (again with the DPT support). For the future plans, see the section on time schedule.

At this stage, TAC project includes:
- Linac-ring type charm factory
- Synchrotron light source based on positron ring
- Free electron laser based on electron linac
- GeV scale proton accelerator
- TAC-Test Facility.

The schematic view of the factory and light sources part of the Turkic Accelerator Complex is given in Fig. 1.

## TAC CHARM FACTORY

Up to now, we have analyzed linac-ring type $\phi$, charm and $\tau$ factory options. In principle $L = 10^{34} \text{cm}^{-2}\text{s}^{-1}$ can be achieved for all three options. Concerning $\phi$ factory option, existing DA$\phi$NE $\phi$ factory has nominal $L = 5\times 10^{32} \text{cm}^{-2}\text{s}^{-1}$ and possible upgrades to higher luminosities are under consideration [7]. Therefore, physics search potential for the $\phi$ factory will be essentially exhausted before TAC commissioning. Concerning $\tau$ factory option, whereas $e^+e^- \to \tau^+\tau^-$ cross-section achieves to maximum value at $\sqrt{s} = 4.2\,GeV$, this advantage is dissipated with success of B-factories which has luminosity of $10^{34} \text{cm}^{-2}\text{s}^{-1}$ already. Moreover super B-factories with $L = 10^{36} \text{cm}^{-2}\text{s}^{-1}$ are intensively discussed [8].

For these reasons, we inclined towards charm factory option. The center of mass energy is fixed by the mass of $\psi(3770)$ resonance. Existing CLEO-c [9] works with $L = 10^{32} \text{cm}^{-2}\text{s}^{-1}$. The BEP charm factory proposal [10] has design luminosity of $10^{33} \text{cm}^{-2}\text{s}^{-1}$. Therefore, TAC charm factory with $L = 10^{34}\,cm^{-2}s^{-1}$, planned to work in mid 2010's, will contribute charm physics greatly. Differing from K and B mesons, where possible new physics manifest itself as a deviation from standard model background, D mesons has negligible standard model background. The main parameter set for TAC charm factory is presented in Table 1. The restriction on luminosity coming from linac beam power can be relaxed by using of energy recovery linac. This topic is under study.

Table 1: Tentative parameters of TAC charm factory

| Parameter | e⁻-linac | e⁺-ring |
|---|---|---|
| Energy, GeV | 1.00 | 3.56 |
| Particles per bunch, $10^{10}$ | 0.55 | 11.00 |
| β function at IP, cm | 0.45 | 0.45 |
| Normalized emittance, μm·rad | 6.17 | 22.00 |
| Bunch length, cm | 0.10 | 0.45 |
| Transverse size at IP, μm | 3.76 | 3.76 |
| Beam-beam tune shift | - | 0.056 |
| Collision frequency, MHz | 30 | |
| Luminosity ($H_D$·L) | $1.4\ 10^{34}\ cm^{-2}s^{-1}$ | |

The luminosity spectrum $dL/dW_{cm}$ obtained using GUINEA-PIG simulation program [14] with $\Delta E/E = 10^{-3}$ is plotted in Figure 2. It is seen that center of mass energy spread is well below $\Gamma_{\Psi(3S)} \approx 24$ MeV. Expected number of Ψ(3S) is about $10^9$ per working year ($10^7$ s). Let us remind that $D^+D^-$ and $D^0D^0$ decay modes are dominant channels for Ψ(3S) decays.

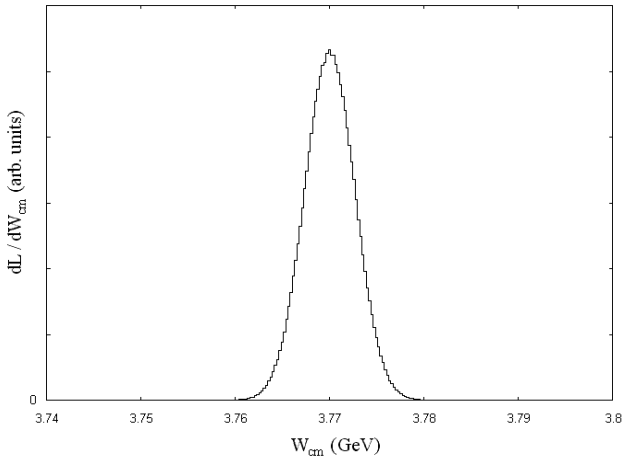

Figure 2: Luminosity spectrum for the TAC charm factory.

## SYNCHROTRON LIGHT SOURCE

Ref. [2] had considered additional positron storage ring dedicated to production of synchrotron radiation. Because of beam-beam tune shift restriction, the emittance of colliding beams in standard (ring-ring) type colliders inevitably should be chosen to be relatively large to obtain high luminosity:

$$L = f_c \frac{4\pi\gamma_p\gamma_e\Delta Q_p\Delta Q_e\varepsilon_p}{r_0^2\beta_e^*} \quad (1)$$

where subscript e (p) corresponds to electron (positron). This restricts the performance of synchrotron radiation obtained from insertion devices placed in standard type colliders.

Fortunately, this is not the case for linac-ring type machines. In this case, emittance of the positron beam does not essentially affect luminosity performance of the co llider:

$$L = f_c \frac{\gamma_p\Delta Q_p N_p}{r_0\beta_p^*} \quad (2)$$

Therefore, the emittance of the positron beam can be chosen small enough to behave as a third generation light source in principle. Normalized emittance of the positron beam given in Table 1 corresponds to transverse emitance of 3 nm·rad, which is well below 20 nm·rad (upper limit for third generation SR sources).

Main parameters of TAC SR Facility were reported at EPAC 2000 [5]. Since then, construction of SESAME [1, 11] has begun in Jordan and CANDLE [12] project has been developed in Armenia. For this reason, final decision on the number of insertion devices and beam lines of TAC SR Facility and their specifications will be made depending on realization of SESAME and CANDLE projects as well as on user potential in our region.

Several samples of optical beam lines design and related studies on TAC SR facility can be found in [13].

## FREE ELECTRON LASER

Main linac of the TAC charm factory can be operated separately to obtain FEL as seen from Figure 1. FEL operation is foreseen during the maintenance of the collider. With 1 GeV electron beam wave length of SASE FEL photons is expected to be a few nm. Detailed studies for different electron beam energies were presented at national conferences [13].

## PROTON ACCELERATOR

TAC proton accelerator proposal consists of 100÷300 MeV energy linear pre-accelerator and 1÷5 GeV main ring. The average beam current values for these machines would be ~30 mA and ~0.3 mA, respectively. Proton beams from two different points of the synchrotron will be forwarded to neutron and muon regions, where a wide spectrum of applied research is planned. In muon region, together with fundamental investigations such as test of QED and muonium-antimuonium oscillations, a lot of applied investigations such as High-$T_c$ superconductivity, phase transitions, impurities in semiconductors *et cetera* will be performed using the powerful Muon Spin Resonance (μSR) method. In neutron region investigations in different fields of applied physics, engineering, molecular biology and fundamental physics are planned.

## TAC TEST FACILITY

Before building charm factory to obtain training of the young accelerator physicists and to get experience on accelerator technology on smaller scale, we plan to build infrared free electron laser (IR FEL) on 20÷50 MeV e-linac until 2009. IR FEL thought to work in oscillator mode. With undulator strength parameter K=1 and vertical distance between magnet poles g=3 cm, one can obtain wavelength values of 15 μm (IR FEL1) and 2 μm (IR FEL2) for 10 MeV and 50 MeV e-linac choices, respectively. Three experimental stations are planed to make research on biomedical subjects, semiconductor physics and photo chemical reactions.

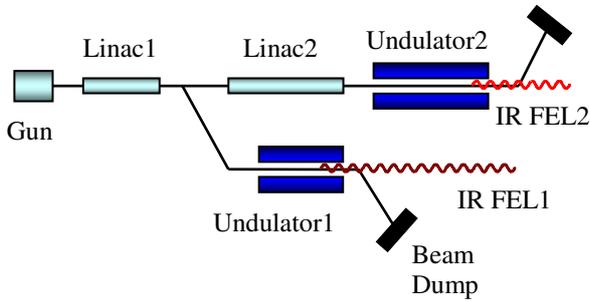

Figure 3: Schematic view of the TAC Test Facility.

## TIME SCHEDULE

Tentative time schedule for the realization of TAC project follows:

2006:
- Completion of the conceptual design report
- Starting technical design study
- Construction of the building for TAC Test Facility

2007:
- Installation of the TAC-TF linac

2008:
- Installation of the TAC-TF infra-red FEL and beam lines with the experimental stations
- Completion of the TAC technical design report

2009:
- Commissioning of TAC-TF
- Governmental decision on approval of TAC project

2015:
- Completion of factory and light source parts of TAC project.

2017:
- Completion of proton accelerator and experimental stations